\documentclass{PoS}
\usepackage{ulem}
\title     {$S$-wave $\pi K$ scattering length from lattice QCD}
\ShortTitle{$S$-wave $\pi K$ scattering length from lattice QCD}
\author{\speaker{Kiyoshi Sasaki} \\
  Department of Physics, Tokyo Institute of Technology, \\
  Tokyo 152-8551, Japan \\
  E-mail: \email{ksasaki@th.phys.titech.ac.jp}
}
\author{Naruhito Ishizuka \\
  Graduate School of Pure and Applied Sciences, University of Tsukuba, \\
  Tsukuba, Ibaraki 305-8571, Japan \\
  and\\
  Center for Computational Science, University of Tsukuba, \\
  Tsukuba, Ibaraki 305-8577, Japan \\
  E-mail: \email{ishizuka@het.ph.tsukuba.ac.jp}
}
\author{Takeshi Yamazaki \\
  Center for Computational Science, University of Tsukuba, \\
  Tsukuba, Ibaraki 305-8577, Japan \\
  E-mail: \email{yamazaki@ccs.tsukuba.ac.jp}
}
\author{Makoto Oka \\
  Department of Physics, Tokyo Institute of Technology, \\
  Tokyo 152-8551, Japan \\
  E-mail: \email{oka@th.phys.titech.ac.jp}
}
\author{for PACS-CS Collaboration}
\abstract{
The $S$-wave $\pi K$ scattering lengths are calculated
for both the isospin 1/2 and 3/2 channels
in the lattice QCD by using the finite size formula.
We perform the calculation with $N_f=2+1$ gauge configurations
generated on $32^3 \times 64$ lattice
using the Iwasaki gauge action 
and nonperturbatively $O(a)$-improved Wilson action at $1/a = 2.17$ GeV.
The quark masses correspond to $m_\pi = 0.30 - 0.70$ GeV.
For $I=1/2$, to separate the contamination from excited states,
we construct a $2 \times 2$ matrix of the time correlation function
and diagonalize it.
Here, we adopt the two kinds of operators, $\bar{s}u$ and $\pi K$.
It is found that the signs of the scattering lengths are in agreement
with experiment, 
namely attraction in $I=1/2$ and repulsion in $I=3/2$.
We investigate the quark-mass dependence of the scattering lengths
and also discuss the limitation of chiral perturbation theory.
}
\FullConference{
  The XXVII International Symposium on Lattice Field Theory - LAT2009 \\
  July 26-31 2009 \\
  Peking University, Beijing, China
}
\begin{document}
%
% @@ ==========================================================================
%
\section  {Introduction}
\label{sec:Introduction}
The scattering length is a key quantity
for understanding the basic properties of the low-energy interaction.
Many lattice calculations of the hadron-hadron scattering lengths
have been reported in past years.
Most of them, however, do not treat
the scattering system with attractive interactions
due to the computational cost.
Handling the attractive interaction
would be indispensable in scattering studies of the future.

Here, we focus on the $S$-wave $\pi K$ system. 
This system has two isospin channels ($I=3/2,\ 1/2$).
The low-energy interaction is repulsive (attractive)
for $I=3/2$ ($I=1/2$).
In addition, existence of a broad resonance is suggested in $I=1/2$.
Until now, three studies in lattice QCD have been reported
\cite{RelatedWorks1, RelatedWorks2, RelatedWorks3}.
The first study was performed by Miao {\it et al.}
\cite{RelatedWorks1}.
They calculated the $I=3/2$ scattering length
within the quenched approximation.
The first calculation with dynamical quarks ($N_f=2+1$) was reported
by the NPLQCD Collaboration
\cite{RelatedWorks2}.
They calculated the $I=3/2$ scattering length for $m_\pi = 0.3-0.6$ GeV.
They further determined the low energy constants
in the $SU(3)$ chiral perturbation theory (ChPT)
and evaluated the $I=1/2$ scattering length by using ChPT.
The first direct calculation on $I=1/2$ has been done
by Nagata {\it et al.}
\cite{RelatedWorks3}.
They, however, used the quenched approximation, 
and ignored the effect of ghost mesons which is nonnegligible in $I=1/2$. 
In addition, their results do not reproduce 
the repulsive interaction for $I=3/2$ at their simulation points. 
The reliability of their calculations remains controversial. 
In conclusion, no satisfactory direct calculation for $I=1/2$ has been carried out.

In the present work, we calculate the $S$-wave $\pi K$ scattering lengths
for both the isospin channels.
We use a technique with a fixed kaon sink operator 
to reduce the computational cost of the calculation of the $\pi K$ for $I=1/2$.
To separate the contamination from excited states for $I=1/2$,
we construct a $2 \times 2$ matrix of the time correlation function
and diagonalize it.
After obtaining the scattering length at each simulation points,
we investigate the quark-mass dependence
and also discuss the limitation of ${\cal O}(p^4)$ $SU(3)$ ChPT.
All calculations of this work have been done
on the super parrarell computers, PACS-CS and T2K-Tsukuba, at the University of Tsukuba.
%
% @@ ==========================================================================
%
\section  {Details of simulation}
\label{sec:Details_of_simulation}
The $S$-wave $\pi K$ scattering length is defined by
\begin{equation} 
  a_0 = \lim_{k\to 0}\ \tan\delta_0(k) / k\ .
\end  {equation}
$k$ is the scattering momentum related to the total energy 
by $E = \sqrt{m_\pi^2 + k^2} + \sqrt{m_K^2 + k^2}$ .
$\delta_0(k)$ is the $S$-wave scattering phase shift 
and can be evaluated by the L\"uscher's finite size formula
\cite{FSM}, 
\begin{equation}
  \left(\ \tan\delta_0(k) / k\ \right)^{-1}
  =
  \frac{1}{\pi L}
  \cdot
  {\cal Z}_{00}\left(1,\frac{k^2}{(2\pi/L)^2}\right)\ ,
\label{eqn:Luschers_formula}
\end  {equation}
where the zeta function ${\cal Z}_{00}$ is an analytic continuation of
\begin{equation}
  {\cal Z}_{00}(s,\bar{n})
  \equiv
  \sum_{\vec{m}\in\mathbf{Z}^3}
  \frac{1}{( \vec{m}^2 - \bar{n})^s} 
\end  {equation}
defined for $\mbox{Re}(s) > 3/2$.
In the case of attractive interaction, 
$k^2$ on the lowest state has a negative value, so $k$ is pure imaginary. 
$\delta_0(k)$ at the unphysical $k$ is no longer physical scattering phase shift. 
${\cal Z}_{00}( 1, k^2/(2\pi/L)^2 )$, however, have a real value even for this case, 
so $\tan\delta_0(k)/k$ obtained by Eq.(\ref{eqn:Luschers_formula}) is also real. 
If $|k^2|$ is enough small, 
we can regard $\tan\delta_0(k)/k$ as the physical scattering length 
at the $\pi k$ threshold ($k=0$) .

For $I=3/2$, one can extract $E$ from the time correlation function
\begin{equation}
  G(t)
  =
  \langle 0|\ 
         K^+ (t_1  )    \pi^+ (t  )          \ 
    ( W_{K^+}(t_0+1) W_{\pi^+}(t_0) )^\dagger\ 
  |0 \rangle
  \cdot \mbox{e}^{m_K(t_1-1)}\ ,
\end  {equation}
where $K^+=\bar{s}\gamma_5 u$, $\pi^+=-\bar{d}\gamma_5 u$, 
and $W_{K^+}$, $W_{\pi^+}$ are the wall-source operators for the corresponding mesons. 
The time slice of the kaon source is shifted from that of the pion source $t_0$ 
to avoid the Fierz mixing of the wall-source operators \cite{FierzMixing}. 
The time slice of the kaon sink operator $t_1$ is fixed as $t_1 \gg t$. 
The exponential factor $\mbox{e}^{m_K(t_1-t)}$ is introduced 
to drop the unnecessary $t$-dependence appearing due to the fixed $t_1$.

For $I=1/2$, 
the existence of the $\kappa$ resonance are suggested in the low energy, 
and then it might be necessary to separate the contamination from the excited states. 
For this purpose, we use the two types of operators 
$\Omega_0$ and $\Omega_1$ ($\overline{\Omega}_0$ and $\overline{\Omega}_1$), 
\begin{eqnarray}
           {\Omega}_0(t  ) &=& \frac{1}{\sqrt{3}} 
                               \left(                K^+ (t_1  )   \pi^0 (t  )
                                      - \sqrt{2}\    K^0 (t_1  )   \pi^+ (t  )
                               \right)
                               \cdot \mbox{e}^{m_K(t_1-t)}\ ,
  \nonumber  \\
           {\Omega}_1(t  ) &=&   \kappa (t    )\ ,
  \nonumber  \\
  \overline{\Omega}_0(t_0) &=& \frac{1}{\sqrt{3}}
                               \left(             W_{K^+}(t_0+1)W_{\pi^0}(t_0)
                                      - \sqrt{2}\ W_{K^0}(t_0+1)W_{\pi^+}(t_0)
                               \right)\ ,
  \nonumber  \\
  \overline{\Omega}_1(t_0) &=& W_\kappa (t_0+1)\ ,
\end  {eqnarray}
where, 
$K^0=\bar{s}\gamma_5 d$, 
$\pi^0=\frac{1}{\sqrt{2}}(\bar{u}\gamma_5 u - \bar{d}\gamma_5 d)$, 
$\kappa=\bar{s}u$, 
and $W_{K^0}$, $W_{\pi^0}$, $W_\kappa$ 
are the wall-source operators for the corresponding mesons. 
The exponential factor in $\Omega_0(t)$ is introduced by the same reason 
as for $I=3/2$. 
We construct the $2\times 2$ matrix of the time correlation function, 
\begin{equation}
  G_{ij}(t)
  =
  \langle 0|\          {\Omega}_i         (t  )\
              \overline{\Omega}_j^\dagger (t_0)\
  |0 \rangle
  \ \ \ \ \ \ \ \ \ \ (\ i,\ j = 0, 1\ )\ ,
\end  {equation}
and with a reference time $t_R$ 
we extract the energy of the ground state by the diagonalization of $G^{-1}(t_R)\ G(t)$ 
\cite{Diagonalization}.

The calculations are carried out 
with $N_f=2+1$ full QCD configurations generated by the PACS-CS Collaboration
\cite{Aoki:2008sm}
using the Iwasaki gauge action at $\beta=1.90$ 
and nonperturbatively ${\cal O}(a)$-improved Wilson quark action 
with $C_{SW}=1.715$ on $32^3 \times 64$ lattice. 
The quark propagators in this work are calculated with the same quark action.
The corresponding lattice cutoff is $1/a=2.176(31)$ GeV ($a=0.0907(13)$ fm) 
and the spatial extent of the lattice is $La=2.902(41)$ fm. 
The quark mass parameters and corresponding hadron masses 
are listed in Table \ref{tbl:mass-param}. 
The Dirichlet (periodic) boundary condition are imposed 
to the temporal direction (spatial directions) in the quark propagators.
The coulomb gauge fixing is employed for the use of the wall source. 
The time slice of the source is 
$t_0=12$ ($t_0+1=13$) for the $\pi$ operator ($K$ and $\kappa$ operators) 
and the fixed sink slice is $t_1=53$ for $K$ operator. 
We adopt $t_R=18$ as the reference time for the diagonalization for $I=1/2$. 
The statistical errors are evaluated 
by the jackknife analysis with a binsize of 250 MD time. 
Here, 
the MD time is the number of trajectories multiplied by the trajectory length $\tau$, 
and $\tau=0.25$ ($\tau=0.5$) for $\kappa_{ud}=0.13770$ (others).

\bigskip
\begin{table}[htbp]
\begin{center}
\begin{tabular}{lllllllll}
\hline
\hline
                                        &
$\kappa_{ud}$    &  $\kappa_s$       &  &
$m_\pi$ [GeV]    &  $m_K$ [GeV]      &  &
$N_\mathrm{conf}$                    &  \\
\hline
                                        &
$0.13770$        &  $0.13640$        &  &
$0.2971(29)$     &  $0.5935\ \ (17)$ &  &
$800$                                &  \\
                                        &
$0.13754$        &  $0.13640$        &  &
$0.4101(27)$     &  $0.6362\ \ (19)$ &  &
$450$                                &  \\
                                        &
$0.13727$        &  $0.13640$        &  &
$0.5707(16)$     &  $0.7135\ \ (14)$ &  &
$400$                                &  \\
                                        &
$0.13700$        &  $0.13640$        &  &
$0.7032(11)$     &  $0.79086(97)$    &  &
$400$                                &  \\
\hline
\hline
\end  {tabular}
\caption{The quark mass parameters and corresponding hadron masses.}
\label{tbl:mass-param}
%
%  Comment by K.S.
%
%    Fit ranges are chosen by the following criterions, 
%
%      1.  plateau in the effective mass plot, 
%      2.  m1 - m2 is zero within statistical errors, where 
%          m1 is the mass obtained from G1(t) with source t=t0+1,
%          m2 is the mass obtained from G2(t) with source t=t0  .
%
%    k_ud = 0.13770 :  [tmin:tmax] = [26:39]  for pion
%                                  = [28:39]  for Kaon
%    k_ud = 0.13754 :  [tmin:tmax] = [26:39]  for pion
%                                  = [32:39]  for Kaon
%    k_ud = 0.13727 :  [tmin:tmax] = [36:46]  for pion
%                                  = [36:46]  for Kaon
%    k_ud = 0.13700 :  [tmin:tmax] = [34:46]  for pion
%                                  = [34:46]  for Kaon
%
\end  {center}
\end  {table}
%
% @@ ========================================================================
%
\section{ Numerical results }
\begin{figure}[htbp]
\begin{center}
\includegraphics[width=155mm]{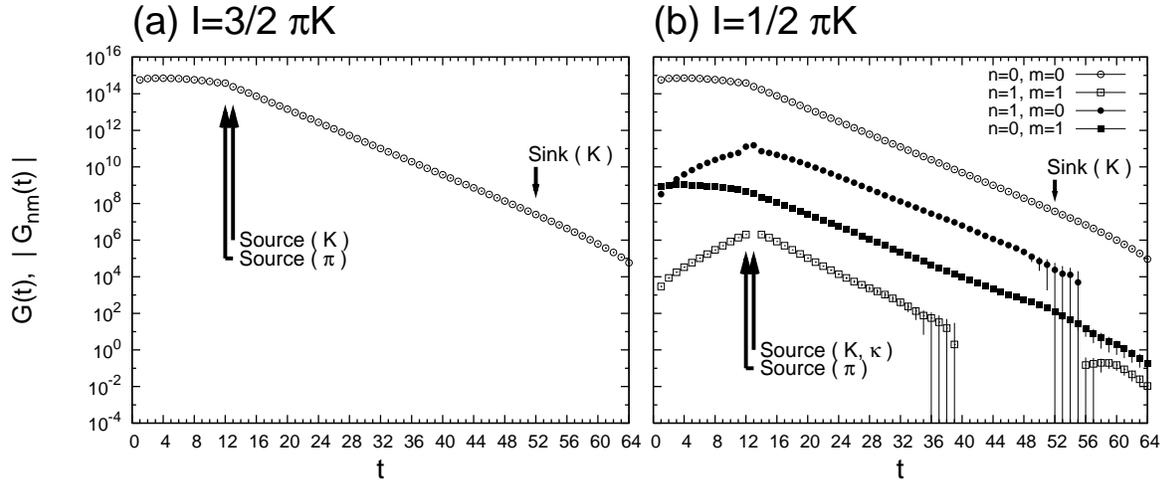}
\caption{(a) $G(t)$ for $I=3/2$ and (b) $|G_{ij}(t)|$ 
         for $I=3/2$ at $m_\pi\simeq 0.30$ GeV.}
\label{fig:time-corr}
\end  {center}
\end  {figure}

In Fig.~\ref{fig:time-corr}, 
we show the time correlation functions 
for both the channels at $m_\pi\simeq 0.30$ GeV as an example. 
In the right panel, the absolute values of each component $G_{ij}(t)$ in $I=1/2$
are presented.
The open (filled) symbols represent 
the (off-)diagonal elements of $G_{ij}(t)$ 
and the signs are positive (negative). 
We find that the signals of $G(t)$ are very clean even in $I=1/2$.
%
%---------------------------------
%---------------------------------

\begin{figure}[t]
\begin{center}
\includegraphics[width=155mm]{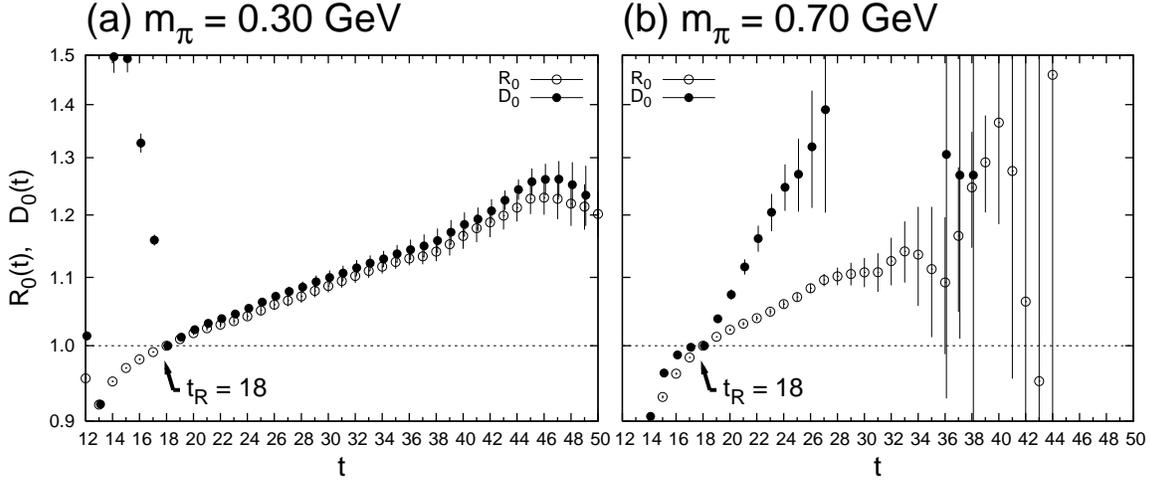}
\caption{$R_0(t)$ and $D_0(t)$ 
         at (a) $m_\pi\simeq 0.30$ GeV and (b) $0.70$ GeV for $I=1/2$. }
\label{fig:corr-ratio}
\end  {center}
\end  {figure}

To see effects of the contaminations from excited states 
for the $I=1/2$ $\pi K$ system, 
we consider the ratios of $G_{00}(t)$ and $\mbox{EV}[\ G^{-1}(t_R)\ G(t)\ ]_0$, 
the lowest eigenvalue of $G^{-1}(t_R)\ G(t)$, to the free $\pi K$ propagator,
\begin{eqnarray}
  R_0(t) &\equiv&
          \frac{G_{00}(t)}{G_{00}(t_R)}
          \cdot
          \left[\ \mbox{e}^{-(m_\pi+m_K)\cdot(t-t_R)}\ \right]^{-1},
  \nonumber  \\
  D_0(t) &\equiv&
          \mbox{EV}\left[\ G^{-1}(t_R)\ G(t)\ \right]_0
          \cdot
          \left[\ \mbox{e}^{-(m_\pi+m_K)\cdot(t-t_R)}\ \right]^{-1}.
\end  {eqnarray}
In the left panel of Fig.~\ref{fig:corr-ratio}, 
$R_0(t)$ (open symbols) and $D_0(t)$ (filled symbols) at $m_\pi\simeq 0.30$ GeV are plotted. 
The difference of the two ratios is small. 
This means that the contamination from the excited states is negligible at this quark mass parameter. 
On the other hand, the right panel of ~\ref{fig:corr-ratio} shows 
that the contamination is not so small at $m_\pi\simeq 0.70$ GeV. 
The diagonalization significantly changes the behavior of the ratio, 
because the $\pi K$-type operator ($\Omega_0$) has a large overlap with the excited states. 
Therefore, separating the contamination is indispensable for the heavy quark masses. 
%
%---------------------------------
%---------------------------------

$k^2$ and $\tan\delta_0(k)/k$ on the lowest state is shown 
in Table.~\ref{tbl:scattering-length}. 
For $I=3/2$ ($I=1/2$), 
$k^2$ is positive (negative), so we confirm the interaction is repulsive (attractive). 
The scattering length is defined as 
the constant term in the $k^2$-expansion of $\tan\delta_0(k)/k$. 
Therefore, $\tan\delta_0(k)/k$ can be regarded as the scattering length 
if $|k^2|$ is small enough to neglect the ${\cal O}(k^2)$ term in the expansion. 
$|k^2|$ for $I=1/2$ is, however, not so small in the heavy quark mass region. 
We especially find an extreme situation in $m_\pi >0.41$ GeV. 
Due to the strong attraction, 
$\tan\delta_0(k)/k$ changes the sign and we get $\tan\delta_0(k)\simeq -i$. 
This fact suggests the appearance of 
an unphysical bound state of the $\pi K$ system in $m_\pi >0.41$ GeV. 
We cannot use $\tan\delta_0(k)/k$ near the bound state 
for the extrapolation toward the $\pi K$ threshold 
because the analytical structure of $\tan\delta_0(k)/k$ is not clear. 
In the following discussion, 
we assume that $|k^2|$ is small enough that 
$\tan\delta_0(k)/k$ reflect information at the $\pi K$ threshold 
for all $m_\pi$ of $I=3/2$ and $m_\pi\le 0.41$ GeV of $I=1/2$, 
and adopt them as the scattering lengths. 
The validity of this assumption must be investigated 
by studying the $L$-dependence of $\tan\delta_0(k)/k$ in the future.

\bigskip
\begin{table}[htbp]
\begin{center}
\begin{tabular}{rcrrrrrrr}
\hline
\hline
                                        &
                 &                      &
\multicolumn{2}{c}{$I=3/2$}             &
                                        &
\multicolumn{2}{c}{$I=1/2$}             &  \\
                                        &
$m_\pi$ [GeV]    & \quad\quad           &
$k^2$ [GeV${}^2$]                       &
$\tan\delta_0(k)/k$ [fm]                &
\quad\quad                              &
$k^2$ [GeV${}^2$]                       &
$\tan\delta_0(k)/k$ [fm]                &  \\
\hline
                                        &
$0.30$           &                      &
$ 0.00326   (30)$                       &
$-0.141     (11)$                       &
                                        &
$-0.00596   (71)$                       &
$ 0.431     (76)$                       &  \\
                                        &
$0.41$           &                      &
$ 0.00404   (82)$                       &
$-0.170     (29)$                       &
                                        &
$-0.0100\ \ (25)$                       &
$ 1.11\ \   (75)$                       &  \\
                                        &
$0.57$           &                      &
$ 0.0029\ \ (10)$                       &
$-0.130     (38)$                       &
                                        &
$-0.0162\ \ (32)$                       &
$-13.\ \ \ \ (120)$                     &  \\
                                        &
$0.70$           &                      &
$ 0.00328   (39)$                       &
$-0.142     (15)$                       &
                                        &
$-0.060\ \ \ \ (12)$                    &
$-\ \ 0.87\ \ (11)$                     &  \\
\hline
\hline
\end  {tabular}
\caption{$k^2$ and $\tan\delta_0(k)/k$ on the lowest state for $I=3/2$ and $I=1/2$.}
\label{tbl:scattering-length}
%
%------------------------------------------------------------------
%
%  Comment by KS.
%
%    t_min is decided to avoid the effect from the excited states. 
%    t_max is decided enough away from t1 ( = 53 ). 
%    However, diag. becomes unstable in t > 30-31 for heavy m_pi.
%    Adopted tmax = 30-31 for heavy m_pi in I=1/2.
%
%    The fitting range is as follows, 
%
%    k_ud = 0.13770 :  [tmin:tmax] = [26:39]  for pipi (I=2  )
%                                  = [26:39]  for piK  (I=3/2)
%                                    [22:39]  for piK  (I=1/2)
%    k_ud = 0.13754 :  [tmin:tmax] = [30:39]  for pipi (I=2  )
%                                  = [31:39]  for piK  (I=3/2)
%                                    [21:31]  for piK  (I=1/2)
%    k_ud = 0.13727 :  [tmin:tmax] = [30:46]  for pipi (I=2  )
%                                  = [30:44]  for piK  (I=3/2)
%                                  = [21:30]  for piK  (I=1/2)
%    k_ud = 0.13700 :  [tmin:tmax] = [34:43]  for pipi (I=2  )
%                                  = [34:46]  for piK  (I=3/2)
%                                  = [20:30]  for piK  (I=1/2)
%
%------------------------------------------------------------------
%
\end  {center}
\end  {table}

We extrapolate the scattering lengths toward the physical point. 
For this purpose, 
we employ the formula predicted by ${\cal O}(p^4)$ $SU(3)$ ChPT
\cite{Bernard:1990kw}. 
To improve the ChPT fit, we also include the data of $\pi\pi (I=2)$, 
which is same scattering system as $\pi K (I=3/2)$ 
except for the replacement of the light and strange quarks. 
In the ${\cal O}(p^4)$ $SU(3)$ ChPT, $a_0$ can be described as 
\begin{eqnarray}
  a_0^{(\pi\pi,I=2)}
  & = &
  \frac{m_\pi}{16\pi F^2}
  \left[\ 
  - 1 
  + \frac{16}{F^2}
    \left[\ 
                     m_\pi^2            \cdot L  (\mu) 
      + \frac{1}{2} (m_\pi^2 + 2 m_K^2 )\cdot L_4(\mu) 
      + \chi^{(\pi\pi,I=2)}\ 
    \right]\ 
  \right]\ ,
  \nonumber  \\
  a_0^{(\pi K,I=3/2)}
  & = &
  \frac{\mu_{\pi K}}{8\pi F^2}
  \left[\ 
  - 1
  + \frac{16}{F^2}
    \left[\ 
                     m_\pi       m_K    \cdot L  (\mu)
      + \frac{1}{2} (m_\pi^2 + 2 m_K^2 )\cdot L_4(\mu)
      + \chi^{(\pi K,I=3/2)}\ 
    \right]\ 
  \right]\ ,
  \nonumber  \\
  a_0^{(\pi K,I=1/2)}
  & = &
  \frac{\mu_{\pi K}}{8\pi F^2}
  \left[\ 
    2
  + \frac{16}{F^2}
    \left[\ 
                      m_\pi       m_K    \cdot L  (\mu)
      -             ( m_\pi^2 + 2 m_K^2 )\cdot L_4(\mu)
      + \chi^{(\pi K,I=1/2)}\ 
    \right]\ 
  \right]\ ,
\end  {eqnarray}
where 
$\mu_{\pi K}$ is the reduced mass of $\pi K$, 
$F$ is the decay constant in the chiral limit, 
and $L_4$, $L \equiv 2L_1 + 2L_2 + L_3 - 2L_4 - L_5/2 + 2L_6 + L_8$ are 
the low energy constants defined in Ref.\cite{Gasser:1984gg} at scale $\mu$. 
$\chi^{( \pi \pi, I=2   )
      ,( \pi K  , I=3/2 )
      ,( \pi K  , I=1/2 )}(\mu,m_\pi,m_K)$ 
are known functions with chiral logarithm terms 
and the explicit forms can be found in the above references.

\begin{figure}[htbp]
\begin{center}
\includegraphics[width=155mm]{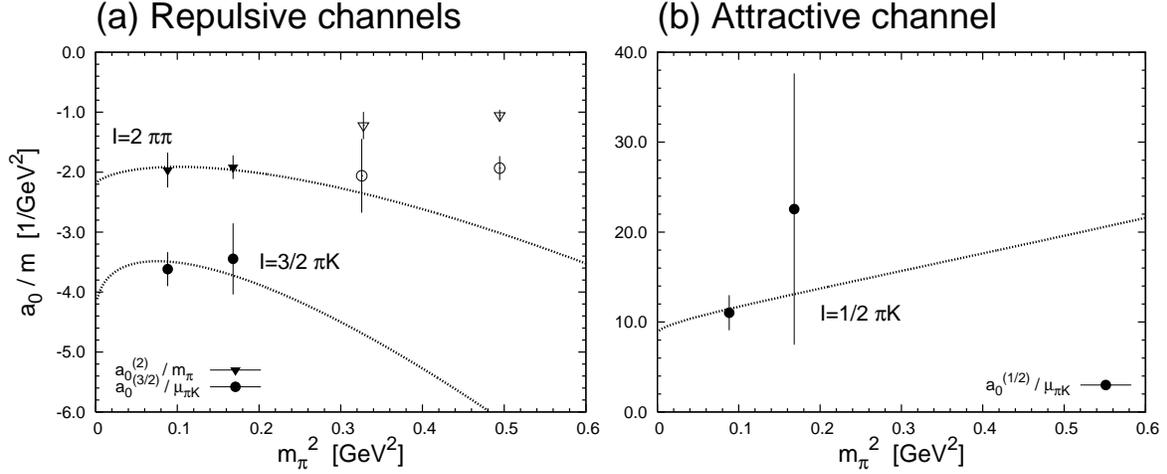}
\caption{$m_\pi^2$-dependence of $a_0/(m_\pi, \mu_{\pi K})$ 
         and fit curves by the ${\cal O}(p^4)$ $SU(3)$ ChPT.}
\label{fig:SU3_ChPT-fit}
\end  {center}
\end  {figure}

The fitting results of 
$a_0^{( \pi \pi, I=2  )}/m_\pi$,\ 
$a_0^{( \pi K  , I=3/2)}/\mu_{\pi K}$\ and\ 
$a_0^{( \pi K  , I=1/2)}/\mu_{\pi K}$ 
are plotted as a function of $m_\pi^2$ in Fig.~\ref{fig:SU3_ChPT-fit}. 
The filled symbols represent the data used in the fit. 
The fit with the data in $m_\pi\ge 0.57$ GeV 
significantly increases $\chi^2/N_\mathrm{df}$, 
so we only use the data in $m_\pi\le 0.41$ GeV. 
The dotted lines are the fitting lines. 
The fit parameters and 
$m_\pi a_0$ at the physical point ($m_\pi=0.140$ GeV, $m_K=0.494$ GeV)
are also listed in Table~\ref{tbl:SU3_ChPT-fit}, 
where the renormalization scale is set to $\mu=0.770$ GeV. 
The numerical results in $m_\pi\le 0.41$ GeV are described 
by the ChPT within the statistical errors. 
$10^3\cdot L_4$ is consistent with $-0.04(10)$ evaluated by PACS-CS Collaboration \cite{Aoki:2008sm}. 
The scattering lengths for $\pi\pi (I=2)$ and $\pi K(I=3/2)$ 
are consistent with the previous lattice studies \cite{RelatedWorks2,Beane:2007xs}. 
Our measured scattering length for $\pi K (I=1/2)$ is also consistent with 
the value evaluated from the data of $\pi K (I=3/2)$ by using ChPT \cite{Beane:2007xs}.

Our evaluation on the scattering length, however, has some issues to be solved. 
First, the statistical errors are not so small. 
The simple solution 
is to add calculations with different source points and improve the statistics. 
Second, 
the behavior near the chiral limit is strongly affected by the chiral logarithm term, 
so giving an evaluation without the long chiral extrapolation is desirable. 
For this purpose, the calculations in $m_\pi< 0.30$ GeV are now in progress. 
Third, 
$\tan\delta_0(k)/k$ in the low-momentum limit must be evaluated 
by systematic studies with the different volumes and boundary conditions. 
These are important issues in the future. 
%
%----------------------
%
%----------------------
%
\bigskip
\begin{table}[htbp]
\begin{center}
\begin{tabular}{cccccccc}
\hline
\hline
$\chi^2/N_\mathrm{df}$           &
$F$ [GeV]                        &
$10^3\cdot L$                    &
$10^3\cdot L_4$                  &
$m_\pi a_0^{(\pi\pi, I=2  )}$    &
$m_\pi a_0^{(\pi K , I=3/2)}$    &
$m_\pi a_0^{(\pi K , I=1/2)}$    \\
\hline
$ 0.3      $                     &
$ 0.114(26)$                     &
$ 0.6  (18)$                     &
$-1.4  (19)$                     &
$-0.037(26)$                     &
$-0.051(38)$                     &
$ 0.14 (10)$                     \\
\hline
\hline
\end  {tabular}
\caption{The fit parameters and
         $m_\pi a_0$ at the physical point 
         ($m_\pi=0.140$ GeV, $m_K=0.494$ GeV) .}
\label{tbl:SU3_ChPT-fit}
\end  {center}
\end  {table}

%
% @@ ==================================================================
%
\section{Conclusion}
\label{sec:Conclusion}
Direct lattice QCD computation of 
the $S$-wave scattering length of $\pi K$ ($I=1/2$ and $I=3/2$) systems 
have been performed. 
The results have reproduced the correct signs of the scattering lengths for the first time 
and therefore have confirmed that the interaction is attractive (repulsive) in $I=1/2$ ($I=3/2$). 
We have found that the attraction in the $\pi K (I=1/2)$ system becomes stronger at $m_\pi> 0.41$ GeV, 
and there the sign of the scattering length becomes negative. 
We have compared the $m_\pi$-dependencies of the scattering lengths 
with those predicted by the ${\cal O}(p^4)$ $SU(3)$ ChPT. 
The data of $\pi K$ ($I=1/2$ and $I=3/2$), $\pi\pi(I=2)$ for $m_\pi\le 0.41$ GeV 
have been used in the fit. 
We have confirmed that the numerical results in $m_\pi\le 0.41$ GeV are described by the ChPT. 
Our evaluation on the scattering length, however, has some issues to be solved, 
and those remain as future tasks. 
%
%
% @@ ==================================================================
%

%

\begin{thebibliography}{50}
%
%---------------------------------------------------------
\bibitem{RelatedWorks1}
%--------------------
C.~Miao, X.~i.~Du, G.~w.~Meng and C.~Liu, 
%``Lattice study on kaon pion scattering length in the I = 3/2 channel,''
Phys.\ Lett.\ B {\bf 595} (2004) 400. 
%
%---------------------------------------------------------
\bibitem{RelatedWorks2}
%--------------------
S.~R.~Beane {\it et al.} [NPLQCD Collaboration],
%``Pi-K Scattering in Full QCD with Domain-Wall Valence Quarks,''
Phys.\ Rev.\ D {\bf 74} (2006) 114503.
%
%---------------------------------------------------------
\bibitem{RelatedWorks3}
%--------------------
J.~Nagata, S.~Muroya and A.~Nakamura,
%``Lattice study of K pi scattering in I = 3/2 and 1/2,''
arXiv:0812.1753 [hep-lat].
%
%---------------------------------------------------------
\bibitem{FSM}
%--------------------
M.~L\"uscher,
%``Volume Dependence Of The Energy Spectrum 
%  In Massive Quantum Field Theories.
%  2. Scattering States,''
Commun.\ Math.\ Phys.\ {\bf 105} (1986) 153 ;\ 
%
%--------------------
%``Two Particle States On A Torus 
%  And Their Relation To The Scattering Matrix,''
Nucl.\ Phys.\ B {\bf 354} (1991) 531.
%
%---------------------------------------------------------
\bibitem{FierzMixing}
%--------------------
M.~Fukugita, Y.~Kuramashi, M.~Okawa, H.~Mino and A.~Ukawa,
%``Hadron scattering lengths in lattice QCD,''
Phys.\ Rev.\ D {\bf 52} (1995) 3003.
%
%---------------------------------------------------------
\bibitem{Diagonalization}
%--------------------
M.~L\"uscher and U.~Wolff,
%``How To Calculate The Elastic Scattering Matrix 
%  In Two-Dimensional Quantum
%  Field Theories By Numerical Simulation,''
  Nucl.\ Phys.\ B {\bf 339} (1990) 222.
%
%---------------------------------------------------------
\bibitem{Aoki:2008sm}
S.~Aoki {\it et al.} [PACS-CS Collaboration],
%``2+1 Flavor Lattice QCD toward the Physical Point,''
arXiv:0807.1661 [hep-lat].
%
%---------------------------------------------------------
\bibitem{Bernard:1990kw}
%
V.~Bernard, N.~Kaiser and U.~G.~Meissner,
%``pi K scattering in chiral perturbation theory to one loop,''
Nucl.\ Phys.\ B {\bf 357} (1991) 129. 
%
%---------------------------------------------------------
\bibitem{Gasser:1984gg}
J.~Gasser and H.~Leutwyler,
%``Chiral Perturbation Theory: Expansions In The Mass Of The Strange Quark,''
Nucl.\ Phys.\ B {\bf 250} (1985) 465.
%
%---------------------------------------------------------
\bibitem{Beane:2007xs}
S.~R.~Beane {\it et al.} [NPLQCD Collaboration],
%``Precise Determination of the I=2 pipi Scattering Length from Mixed-Action
%  Lattice QCD,''
Phys.\ Rev.\ D {\bf 77} (2008) 014505.
%
%---------------------------------------------------------
%
\end{thebibliography}
\end{document}